\documentclass[%
 reprint,
showpacs,preprintnumbers,
 amsmath,amssymb,
 aps,
prd,
]{revtex4-1}

\usepackage{graphicx}
\usepackage{dcolumn}
\usepackage{bm}
\usepackage{hyperref}
\usepackage{color}
\usepackage{braket}
\usepackage[utf8]{inputenc}

\DeclareRobustCommand{\eqnr}[1]{Eq.~$\left(\ref{#1}\right)$}
\DeclareRobustCommand{\order}[1]{\mathcal{O}(#1)}

\DeclareRobustCommand{\abs}[1]{\left\lvert #1 \right\rvert }
\DeclareRobustCommand{\cross}{\boldsymbol{\times}}
\DeclareRobustCommand{\curl}[1]{\nabla \cross #1}

\newcommand{\eqnrtwo}[2]{Eqs.~$\left(\ref{#1}\right)$ and $\left(\ref{#2}\right)$}

\newcommand{\Fig}[1]{Fig.~\ref{#1}}

\newcommand{\Sec}[1]{Section \ref{#1}}

\newcommand{\piz}{{\pi^0}}
\newcommand{\mpiz}{m_{\piz}}

\newcommand{\pid}{\pi^0_d}

\newcommand{\ub}{\overline{u}}
\newcommand{\db}{\overline{d}\,}
\newcommand{\vs}{\vec{s}}

\newcommand{\w}{{\mathrm{w}}}
\newcommand{\wb}{{[\mathrm{w}]}}
\newcommand{\rb}[1]{\left(#1\right)}

\newcommand{\Slash}{\!\!\!\!/\,}

\begin{document}

\preprint{ADP-19-25/T1105}

\title{Pion in a uniform background magnetic field with clover fermions}

\author{Ryan Bignell}
 \email{ryan.bignell@adelaide.edu.au}
\author{Waseem Kamleh}
\author{Derek Leinweber}

\affiliation{%
Special Research Centre for the Subatomic Structure of Matter (CSSM),\\
Department of Physics, University of Adelaide, Adelaide, South Australia 5005 Australia
}%



\date{\today}
             
\begin{abstract}

Background field methods provide an important nonperturbative formalism for the determination of
hadronic properties which are complementary to matrix-element calculations.  However, new
challenges are encountered when utilising a fermion action exposed to additive mass
renormalisations.  In this case, the background field can induce an undesired field-dependent
additive mass renormalisation that acts to change the quark mass as the background field is
changed.
For example, in a calculation utilising Wilson fermions in a uniform background magnetic field,
the Wilson term introduced a field-dependent renormalisation to the quark mass which manifests
itself in an unphysical increase of the neutral-pion mass for large magnetic fields.  
Herein, the clover fermion action is studied to determine the extent to which the removal of
$\order{a}$ discretisation errors suppresses the field-dependent changes to the quark mass.  We
illustrate how a careful treatment of nonperturbative improvement is necessary to resolve this
artefact of the Wilson term.  Using the $32^3 \times 64$ dynamical-fermion lattices provided by the
PACS-CS Collaboration we demonstrate how our technique suppresses the unphysical mass
renormalisation over a broad range of magnetic field strengths.
\end{abstract}


\maketitle

\section{Introduction}

The uniform background-field method~\cite{Smit:1986fn,Martinelli:1982cb,Burkardt:1996vb} has a
long history of use in lattice QCD to investigate and calculate quantities such as
hadronic magnetic moments~\cite{Martinelli:1982cb,Tiburzi:2012ks,Bernard:1982yu,Primer:2013pva},
polarisabilities~\cite{Tiburzi:2012ks,Primer:2013pva,Hall:2013dva,Luschevskaya:2014lga,Bignell:2018acn},
and
the spatial distribution of quarks in a magnetic field~\cite{Roberts:2010cz}.
Recently, it has been demonstrated that the background-field method introduces unphysical changes
in the fermion energy when the Wilson quark formulation is used~\cite{Bali:2015vua,Bali:2017ian}.

Bali {\it et al.} highlighted this important problem~\cite{Bali:2015vua,Bali:2017ian} arising from
the Wilson fermion action with the background field method. In particular they determined that in
the free-field limit, that the mass of a Wilson quark is shifted by an amount
$\frac{a}{2}\,\abs{qe\,B}$ to
\begin{align}
  m_{\wb}(B) = m_\w(0) + \frac{a}{2}\,\abs{qe\,B}\,,
  \label{eqn:mqB}
\end{align}
where $a$ is the lattice spacing, $B$ the magnetic field strength, $qe$ is the quark charge, and
$m_\w$ is the Wilson quark mass. Here we have introduced the notation that a subscript label in square brackets represents a quantity that is affected by lattice background field artefacts due to the fermion action, in this case the Wilson quark mass $m_{\wb}(B).$ This notation will be used throughout this work in order to distinguish between energy values with and without the additive background field mass renormalisation respectively (e.g. to distinguish the pion mass $m_{[\pi]}(B)$ and $m_\pi$).

It was shown in Refs.~\cite{Luschevskaya:2014lga,Bali:2015vua} that the
overlap quark formalism~\cite{Ginsparg:1981bj,Neuberger:1997fp} does not suffer from this problem
of field-dependent mass renormalisation. As the overlap action is many times more computationally
expensive than the Wilson action, we turn our focus to the Wilson clover fermion action to
determine its suitability with respect to the additive mass renormalisation arising from the
background field.

The clover fermion action~\cite{Sheikholeslami:1985ij} is designed to remove $\order{a}$ lattice
artefacts arising from the Wilson term. Thus, it is interesting to examine the extent to which the
Wilson fermion artefacts survive in the clover fermion formulation. Herein, the clover fermion is
studied in both the free-field limit and full QCD to determine its efficacy in removing the
unphysical energy change caused by the uniform background magnetic-field.

Pion correlation functions are the natural choice with which to investigate this issue; they are
free of complications associated with the magnetic moment, provide precision at low computational
cost, and offer insight into interesting physics through the difference between neutral and charged
pions.  The neutral pion is of particular interest as it also has no hadronic-level Landau contribution.

The free-field limit allows the unphysical additive mass renormalisation to be studied, without the
complications of QCD. This aids in formulating a solution for this problem using clover
fermions. When full QCD interactions are present, the competing effects of QCD and the background
magnetic-field make isolating and understanding the additive mass renormalisation more
challenging. Moreover, the use of a nonperturbatively improved clover coefficient in a full QCD
calculations further complicates the transition from free-field to full QCD interactions.

Here, the clover fermion action is studied for both the free-field and full-QCD cases to determine the
conditions for the removal of the field-dependent additive quark mass renormalisation.
In \Sec{sec:BFM} a brief overview of the background field method is presented while \Sec{sec:F-Fc}
details the calculations performed in the free-field limit. These confirm the presence of the
additive mass renormalisation and the utility of the clover fermion action in this limit. Full QCD
is considered in \Sec{sec:FQCD} and \Sec{sec:EMClover} details how the clover fermion action can be
tuned to remove the field-dependent additive quark mass renormalisation.  The magnetic
polarisability of the neutral pion is presented in \Sec{sec:MagPol} and conclusions are summarised
in \Sec{sec:Conclusions}.

\section{Background Field Method}
\label{sec:BFM}

In this approach, a constant background magnetic field is introduced
and the eigenstates are examined in the basis of the full Hamiltonian~\cite{Burkardt:1996vb}.  We commence in the continuum where a
minimal electromagnetic coupling is added to form the covariant
derivative
\begin{align}
	D_\mu = \partial_\mu + i\,qe\,A_\mu\,.
\end{align}
Here $qe$ is the electric charge of the fermion field and $A_\mu$ is
the electromagnetic four potential. On the lattice, the equivalent
modification is to multiply the QCD gauge links by an exponential
phase factor
\begin{align}
	U_\mu(x) \rightarrow U_\mu(x)\,e^{i\,a\,qe\,A_\mu(x)} \, .
\end{align}
To generate a uniform magnetic field along the $\hat{z}$ axis in the
continuum, one considers
\begin{subequations}
\begin{align}
	\vec{B} &= \curl{\vec{A}} \, ,  \\
	B_z &= \partial_x\,A_y - \partial_y\,A_x \, .
\end{align}
\end{subequations}
This does not uniquely specify the electromagnetic potential. The
choice commonly selected over the interior of the lattice is $A_x =
-B\,y$ which gives a constant magnetic field of magnitude $B$ in the
$+\hat{z}$ direction. Due to the periodic boundary conditions on the
lattice, we set $A_y = + B\,N_y\,x$ along the boundary in the
$\hat{y}$ dimension to maintain the constant magnetic field across the
boundary.  Now, considering the entirety of the lattice, a
quantisation condition~\cite{Smit:1986fn} for the magnetic field
strength is required
\begin{align}
	qe\,B\,a^2 = \frac{2\pi\,k_B}{N_x\,N_y} \, ,
	\label{eqn:qc}
\end{align}
where $a$ is the lattice spacing, $N_x$, $N_y$ are the spatial
dimensions of the lattice and $k_B$ is an integer specifying the field
strength in terms of the minimum field strength.

In this work the field quanta $k_B$ is in terms of the charge of the
down quark, {\it i.e.}, $q=-1/3$ such that
\begin{align}
	e B = \frac{-6\,\pi\,k}{a^2\,N_x\,N_y} \, .
	\label{eqn:eB}
\end{align}
Hence a field with $k_B=1$ will be in the $-\hat{z}$ direction.

\section{Free-field case}
\label{sec:F-Fc}
The free-field simulation is investigated first; here the quarks
couple to the external magnetic field through their electric charges
but do not experience any QCD effects. The energy of such a charged
particle contains a Landau energy term proportional to the charge of
the particle. In the non-relativistic approximation with
$\vec{B}=B\,\hat{z}$, this energy spectrum is equivalent to that of a
harmonic oscillator
\begin{align}
  E = \left(n + \frac{1}{2}\right)\,\frac{\abs{qe\,B}}{m} \, ,\quad n=0,1,\dots \, .
\end{align}
The relativistic generalisation of the Landau energy applies to each
fermion responding to the field. For a free quark, the energy is
\begin{align}
  E^2(B) = m^2 + \left(2\,n+1\right)\,\abs{qe\,B} + p_z^2 + 2\,\vs\cdot qe\,\vec B \, .
\end{align}
Here $p_z$ is the quark momentum in the $\hat{z}$ direction,
$\abs{\vs} = 1/2$ and the quark has charge $qe$.

\subsection{Wilson Fermions}

Wilson fermions will
have an additional energy term according to \eqnr{eqn:mqB}, and thus a
free-field energy
\begin{align}
  E_{\wb}^2(B) = &\left(m_\w + \frac{a}{2}\,\abs{qe\,B}\right)^2 \nonumber \\
  &+ \left(2\,n+1\right)\,\abs{qe\,B} + p_z^2 + 2\,\vs\cdot qe\,\vec B\,,
  \label{eqn:EWilson}
\end{align}
as the Wilson term is the discretised lattice Laplacian which also
describes the lowest lying Landau level~\cite{Bignell:2018acn}.

As $a$ is the lattice spacing, this additional term is absent in the
continuum limit. The additional term in \eqnr{eqn:EWilson} is a
lattice artefact identified with field-strength dependent additive
quark mass renormalisation.

To demonstrate the presence of this additive mass term, the free-field
pion mass is calculated.  Here we consider both the charged pion
energy, $E_{\pi^\pm}$ and the neutral connected pion energy
$E_{\pi^0_{u/d}}(B)$ and focus on the lowest lying states.  Standard
pseudoscalar interpolating fields $\chi = \bar{q}\,\gamma_5\,q$ are
considered, where the quark flavours are either $u \ub$ or $d \db$
corresponding to $\pi^0_u$ and $\pi^0_d$.

For a neutral pion with quark content $u\ub$ or $d\db$ it is possible
to have the the spin-dependent term of \eqnr{eqn:EWilson},
$2\,\vs\cdot qe\,\vec B$ cancel the Landau energy term
$\left(2\,n+1\right)\,\abs{qe\,B}$ for $n=0$.  This cancellation of
terms occurs for both the quark and the anti-quark.
Consider for example, the $u\ub$ pseudoscalar.  As a spin-zero state,
the two quarks have opposite spin orientations. Similarly, the quark
and antiquark have opposite electric charges.  If the $u$ quark is
spin down and the $\ub$ is spin up, the terms cancel.
For $p_z = 0$, one has the lowest lying state with energy
\begin{align}
  E_{[\piz]}(B) &= E_{[u]}(B) + E_{[{\bar u}]}(B) \nonumber \\
  &= m_{u}   + \frac{a}{2}\,\abs{q_{u}  e B}
   + m_{\bar u} + \frac{a}{2}\,\abs{q_{\bar u}e B} \nonumber \\
  &= \mpiz + a\,\abs{q_{u}e B} \, .
  \label{eqn:FFPiN}
\end{align}

For a charged pion the spin-dependent term cannot cancel the Landau
term for both the quark and antiquark sectors.  The lowest energy
state is realised when the terms cancel for the quark flavour with the
largest magnitude of electric charge.

Consider for example, the $\pi^+$ meson composed of $u\db$.  Here the
lowest energy state is realised when the $u$ quark is spin down,
enabling cancellation of the spin-dependent and Landau terms.  For $p_z
= 0$, the charged pion will hence have energy
\begin{align}  
  E_{[{\pi^+}]}(B) &= E_{[u]}(B) + E_{[{\bar d}\,]}(B) \nonumber \\
  &= \phantom{+}m_{u} + \frac{a}{2}\,\abs{q_{u}e B} \nonumber \\
  &\phantom{=}+ \sqrt{ \left(m_{\bar d} + \frac{a}{2}\,\abs{q_{\bar d}e B}\right)^2 + 2\,\abs{q_{\bar d}e B}} \, .
   \label{eqn:FFPiPlus}
\end{align}
Note that as QCD interactions are not present, no energy is required
to displace the quarks from each other and thus the magnetic
polarisability vanishes.

In the  absence of the Wilson background-field additive mass
renormalisation, the charged pion energy becomes
\begin{align}  
  E_{\pi^+}(B) &= E_u(B) + E_{\bar d}(B) \nonumber \\
  &= \phantom{+}m_{u} + \sqrt{ m_{\bar d}^2 + 2\,\abs{q_{\bar d}e B}} \, .
   \label{eqn:FFPiPlusClover}
\end{align}

\subsection{Clover correction}
\label{sec:FFEMClover}

Clover fermions are designed to remove
$\order{a}$ artefacts arising in the Wilson action. Thus, the focus
of this investigation is to determine whether the clover fermion
action removes the $\order{a}$ field strength dependent additive mass
renormalisation due to the Wilson term. 

The clover fermion matrix is given by
\begin{equation}
  D_{\rm cl} = \nabla\Slash + \frac{a}{2}\Delta - a\,c_{\textrm{cl}}\, \sigma\cdot F + m\,,
  \label{eqn:clover}
\end{equation}
where $a$ is the lattice spacing, $m$ is the bare quark mass, $\nabla$ is the covariant central finite difference operator, $\Delta$ is the lattice Laplacian or Wilson term, and $c_{\textrm{cl}}$ is the coefficient of the clover term. We define
\begin{equation}
  \sigma\cdot F = \sum_{\mu < \nu} \sigma_{\mu\nu}F_{\mu\nu}\,.
\end{equation}
as the clover term (note the sum is restricted to $\mu < \nu$ to avoid double counting), where $\sigma_{\mu\nu} = \frac{i}{2}[\gamma_\mu,\gamma_\nu]$ and $F_{\mu\nu}$ is the clover discretisation of the lattice field strength tensor. In general, for QCD+QED calculations we need to consider the chromomagnetic and electromagnetic field strength contributions separately~\cite{Bali:2017ian},
\begin{equation}
F_{\mu\nu} = F^{\textrm{qcd}}_{\mu\nu} + F^{\textrm{em}}_{\mu\nu}\,.
\end{equation}
In the free-field case, all the QCD links are set to 1 so the QCD field strength $F^{\textrm{qcd}}_{\mu\nu} = 0.$ Here we consider only the electromagnetic field strength, and attempt to determine the appropriate value for the free parameter $c_{\textrm{cl}},$ which is defined to be the coefficient of the electromagnetic clover term.
Setting
\begin{multline}
C_{\mu\nu}(x) = \frac{1}{4}(P_{\mu,\nu}(x)+P_{\nu,-\mu}(x)\\+P_{-\nu,\mu}(x)+P_{-\mu,-\nu}(x))\,,
\end{multline}
to be the average of the four elementary plaquettes, it is easy to show that the electromagnetic field strength tensor is constant in a uniform background field
\begin{align}
  F_{\mu\nu} &= \frac{1}{2i}(C_{\mu\nu} - C_{\mu\nu}^\dagger)\, \nonumber \\
  \!\implies F_{12}&= \frac{1}{2i}(e^{+ia^2qB} - e^{-ia^2qB}) \nonumber \\
  &= \sin a^2qB\,.
\end{align}
For convenience we define the lattice magnetic field strength as
\begin{equation}
  B_L = \sin a^2qB.
\end{equation}
For a magnetic field in the $z$ direction then the only non-zero field strength tensor entry is $F_{12} = B_L$ such that $\sigma\cdot F$ becomes diagonal with entries $\mp B_L/2.$ 
The Wilson term is simply the lattice Laplacian, which effectively describes a scalar particle. In the free field case, this causes a shift in the critical mass given by
\begin{equation}
  am_c(B) \simeq am_c - a^2|q B|/2\,,
\end{equation}
such that the free field mass is adjusted by
\begin{equation}
  am(B) \simeq am(0) + a^2|q B|/2\,.
\end{equation}
Hence, in a uniform magnetic field there is a corresponding Landau energy $\sim a^2 |qB|/2.$


The magnetic part of the clover term for $O(a)$ improvement of the Wilson matrix is given by
\begin{equation}
  -c_{\textrm{cl}} \sum_{\mu<\nu} \sigma_{\mu\nu}F_{\mu\nu} =  -\frac{c_{\textrm{cl}}}{2}\begin{pmatrix} +B_L & 0 & 0 & 0 \\ 0 & -B_L & 0 & 0 \\ 0 & 0 & +B_L & 0 \\ 0 & 0 & 0 & -B_L \end{pmatrix}\,.
\end{equation}
This operator commutes with the lattice Laplacian, and as both are Hermitian it is possible to write down a shared eigenvector basis. The lowest eigenvalue for the clover term in a uniform background field is
\begin{align}
  \lambda_{\rm min} &= -c_{\textrm{cl}}|B_L|/2  \nonumber \\
  &= -c_{\textrm{cl}}|\sin a^2qB|/2  \nonumber \\
  &\simeq -c_{\textrm{cl}}\,a^2|qB|/2\,.
\end{align}
It is clear then that for small field strengths, if we set $c_{\textrm{cl}} = 1$ to the tree-level value, the clover term will cancel the Landau shift induced by the Wilson term.

\begin{figure}[t!]
	\begin{center}
	\includegraphics[width=\columnwidth]{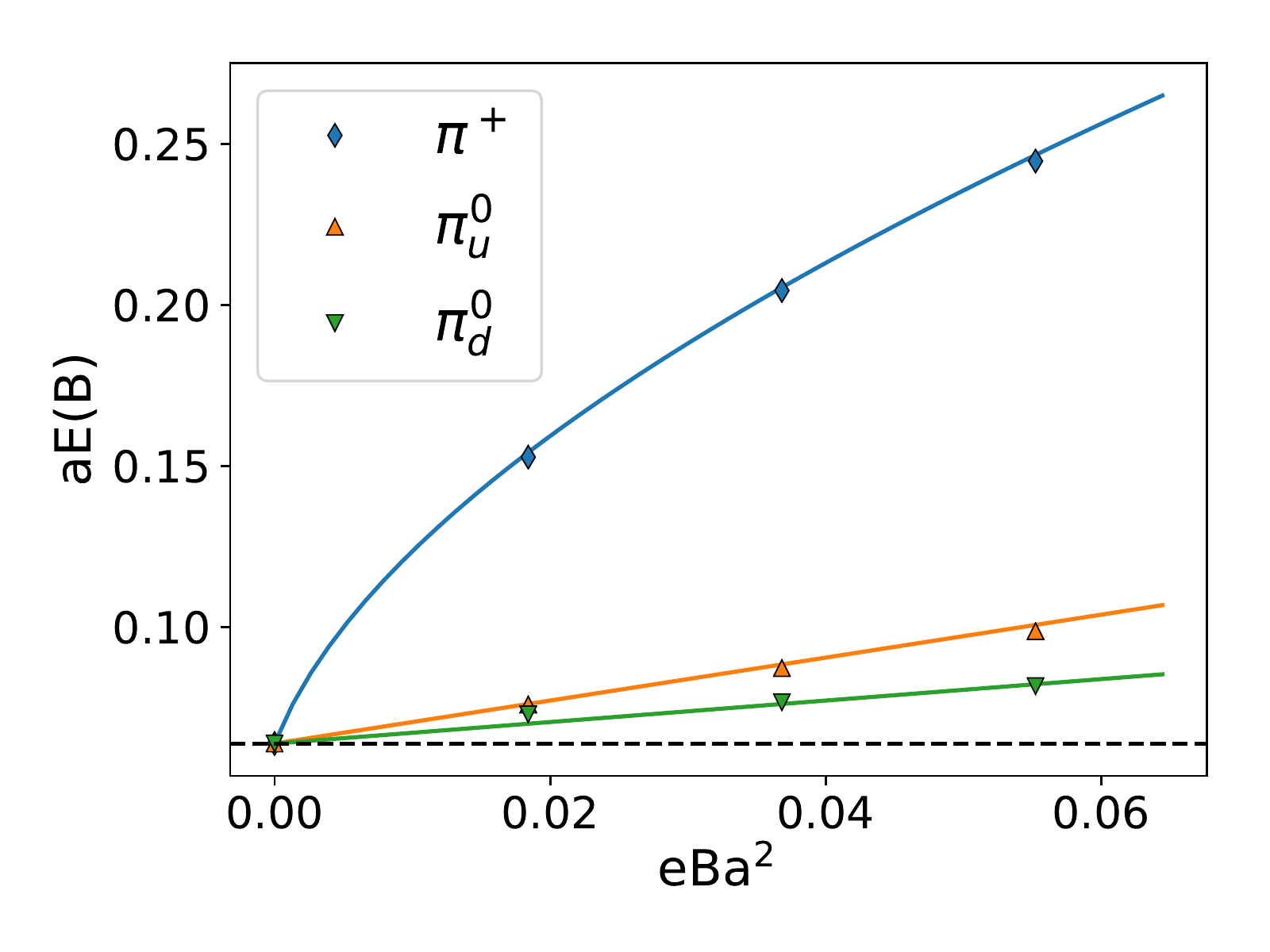}
        \end{center}
	\caption{Pion energies from Wilson-fermion correlation
          functions as a function of background magnetic field
          strength.  The coloured curves are the expected energies for
          Wilson fermions based on \eqnrtwo{eqn:FFPiN}{eqn:FFPiPlus}.}
	\label{fig:FF_Wilson_action}
\end{figure}

\begin{figure}[t!]
	\begin{center}
	\includegraphics[width=\columnwidth]{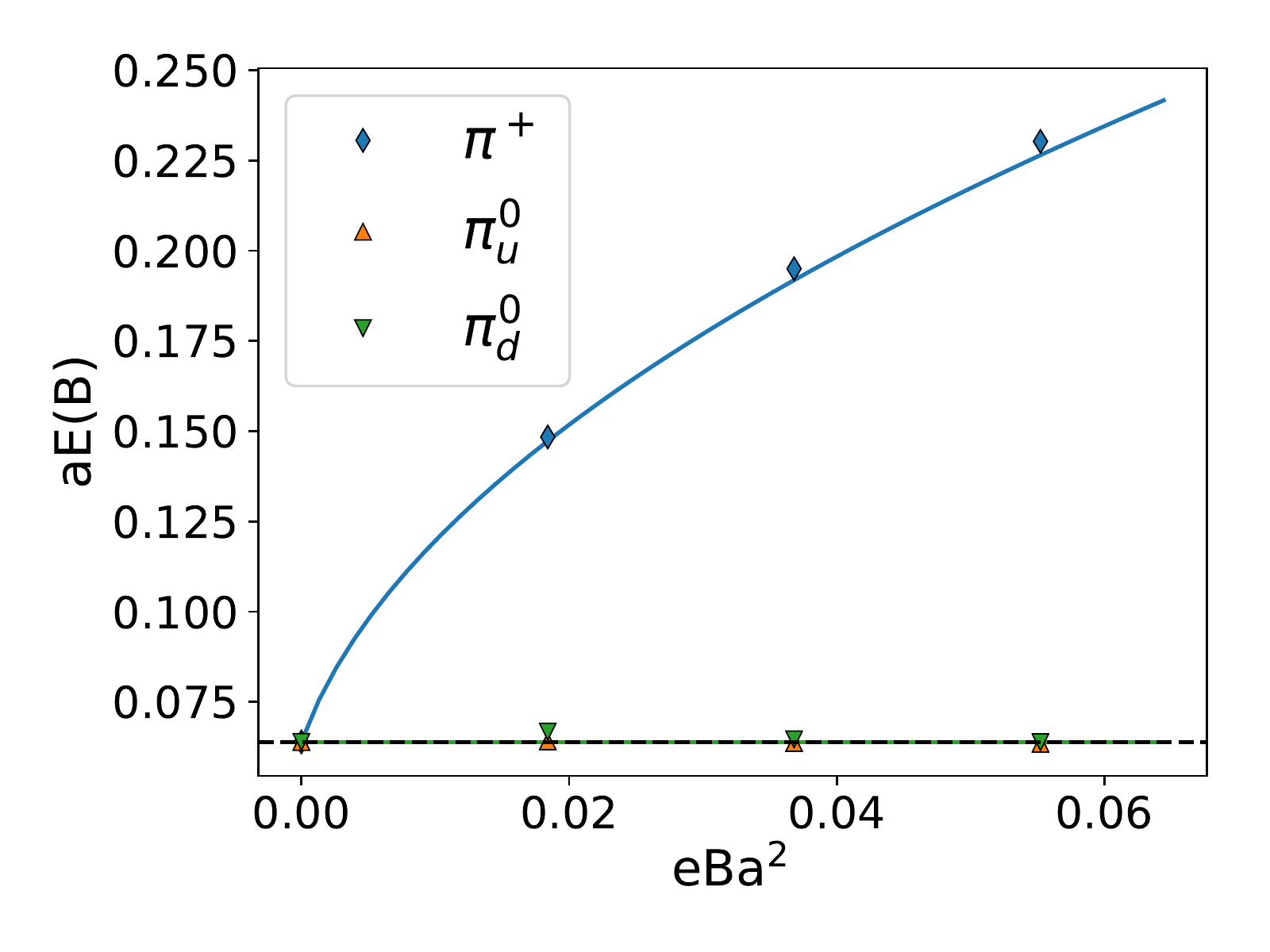}
        \end{center}
	\caption{Pion energies from clover-fermion correlation
          functions as a function of background magnetic field
          strength. The coloured lines are the expected energies in
          the absence of the Wilson background-field additive mass
          renormalisation, constant for the neutral pions and
          Eq.~(\ref{eqn:FFPiPlusClover}) for the charged pion.}
	\label{fig:FF_Clover_action}
\end{figure}

\subsection{Tree-level results}

Numerical results for the pion energies in a background magnetic field
are shown in \Fig{fig:FF_Wilson_action} for Wilson fermions and
\Fig{fig:FF_Clover_action} for clover fermions. Here we have selected a hopping parameter value of $\kappa = 0.12400$, relative to the bare critical hopping parameter of $\kappa_{cr} = 1/8\,$; in anticipation of exploring full QCD where the PACS-CS Sommer scale provides $a = 0.0951$ fm. This kappa value corresponds to a pion mass of $140$ MeV. 
These calculations are performed on a lattice with anti-periodic boundary conditions in
the time direction and fits accommodate both the forward and backward
propagating states.

It is clear from \Fig{fig:FF_Wilson_action} that the neutral pion
energy closely follows that described by \eqnr{eqn:FFPiN}. The
additive mass renormalisation due to the Wilson term is hence
correctly described by \eqnrtwo{eqn:mqB}{eqn:FFPiN}. The charged
$\pi^+$ energy using Wilson fermions agrees with the analytic energy
of \eqnr{eqn:FFPiPlus}, further validating the additive mass
renormalisation of \eqnr{eqn:mqB}.

The results shown in \Fig{fig:FF_Clover_action} for the clover fermion action verify our calculation for the tree level coefficient correction.
The energies of the two neutral pions, $\pid$ and $\pi^0_u$ depicted
 for clover-improved fermions do not show
this additive mass renormalisation; they do not change as a function
of field strength.  Similarly, the charged pion results agree with
Eq.~(\ref{eqn:FFPiPlusClover}) for the pion energy in the absence of
the Wilson background-field additive mass renormalisation.  Thus the
clover term effectively removes the ${\cal O}(a)$ additive mass
renormalisation of the Wilson term in the background magnetic field.

\section{QCD Case}
\label{sec:FQCD}

We have demonstrated that the addition of the tree level clover term
in the case of the electromagnetic $U(1)$ field provides a correction
that removes the background field dependent additive mass
renormalisation induced by the Wilson term.  The next
question that arises is whether this correction survives once QCD
interactions are turned on, and what (if any) adjustments are needed
to the clover coefficient.

To study this, we allow for the QCD and electromagnetic field
strengths to have different clover coefficients,
\begin{equation}
  c_{\textrm{cl}}F_{\mu\nu} \rightarrow c_{\textrm{sw}}F^{\textrm{qcd}}_{\mu\nu} + c_{\textrm{em}}F^{\textrm{em}}_{\mu\nu}\,.
  \label{eq:emclover}
\end{equation}
Here $c_{\textrm{sw}}$ is the coefficient of the traditional Sheikholeslami-Wohlert improvement term arising from the $SU(3)$ QCD gauge field interactions, and $c_{\textrm{em}}$ is the coefficient of the clover term induced by the $U(1)$ electromagnetic gauge field.

We know that the $c_{\textrm{sw}}$ coefficient is renormalised by
QCD. In particular, here we will be studyng clover fermions with a
non-perturbatively improved (QCD) coefficient. To determine the
appropriate value for $c_{\textrm{em}}$ in this instance, we will
first consider the `naive' case, and set $c_{\textrm{em}} =
c_{\textrm{sw}}$ such that the clover terms from the QCD and
electromagnetic field strengths are treated in a uniform way. Having
done this, we will look for the presence of magnetic field strength
dependent artefacts in the pion mass, as was done for the QCD
free-field case.

Once QCD interactions are present, the pion possesses an internal
structure and mass which is dependent on the mass of the quarks in a
more subtle way~\cite{GellMann:1968rz,Gasser:1982ap} than in
\eqnrtwo{eqn:FFPiN}{eqn:FFPiPlus}. There is now a complex interplay
between the background field and QCD effects~\cite{Bruckmann:2017pft,Bignell:2018acn}.

The energy of a relativistic pion with mass $m_\pi$ and charge $qe$ in
a magnetic field orientated in the $\hat{z}$-direction is
\begin{align}
  E^2_{\pi,n}(B) = &m_\pi^2 + \left(2\,n+1\right)\,\abs{qe\,B} + p_z^2 \nonumber \\
  -&4\pi\,m_\pi\,\beta_\pi\,B^2 + \order{B^3}\,,
	\label{eqn:EBpi}
\end{align}
where $\beta_\pi$ is the magnetic polarisability, which is
charge-state dependent. It is necessary to use this fully relativistic
form of the energy for the pion rather than the Taylor expansion
common in the literature~\cite{Martinelli:1982cb,Bernard:1982yu,Tiburzi:2012ks,Primer:2013pva,Bignell:2018acn}
as $2m/(E+m)$ differs substantially from one for the largest field
strength~\cite{Bignell:2018acn}. At $m_\pi = 296$ MeV this difference
is as much as $22\%$ and is significant when working with the
increased precision of pion correlation functions.

\subsection{Simulation Details}

This work considers the $2+1$ flavour dynamical gauge configurations
provided by the PACS-CS~\cite{Aoki:2008sm} collaboration via the ILDG~\cite{Beckett:2009cb}. These configurations are founded on a
nonperturbatively improved clover fermion action and Iwasaki gauge
action~\cite{Iwasaki:2011np}. Two values of the light quark hopping
parameter $\kappa_{ud} = 0.13754,\,0.13770$ are considered
corresponding to pion masses $m_\pi = 411$ and $296$ MeV~\cite{Aoki:2008sm}.  The lattice spacing for each mass is set using
the Sommer scale~\cite{Aoki:2008sm} with $r_0 = 0.49$ fm. The lattice
volume is $L^3 \times T=32^3 \times 64$ and the ensemble sizes are 449
and 400 configurations respectively.  Unless otherwise mentioned, we
focus on results from the lighter ensemble with $\kappa_{ud} =
0.13770$ and $m_\pi = 296$ MeV.

Two point correlation functions at three distinct non-zero background
magnetic field strengths are calculated. To do this, propagators at
ten non-zero field strengths are created.  With reference to
Eq.~(\ref{eqn:eB}), the field strengths considered have $k_B = \pm 1, \,
\pm 2,\, \pm 3,\,\pm 4$, and $\pm 6$ in Eq.~(\ref{eqn:eB}). These give
physical field strengths of $eB = \pm 0.087$, $\pm 0.174$, $\pm
0.261$, $\pm 0.348$, and $\pm 0.522$ GeV$^2$. Correlation functions
are averaged over both positive and negative field strengths during
analysis to provide an improved unbiased estimator.

We note that these configurations are electro-quenched; the field
exists only for the valence quarks of the hadron. To include the
background field at configuration generation time is possible~\cite{Fiebig:1988en} but requires a separate hybrid-Monte-Carlo
calculation for each field strength, which is prohibitively
expensive.  Fortunately, these effects are not relevant to the current
investigation.

Three-dimensional spatial Gaussian smearing utilising stout-smeared
links is applied at the source and a point sink is considered. This
ensures that the pion ground state is well represented.  Standard
pseudoscalar interpolating fields $\chi = \bar{q}\,\gamma_5\,q$ are
used, where the quark flavours are either $u \ub$ or $d \db$
corresponding to $\pi^0_u$ and $\pi^0_d$.

\subsection{Energy Shifts for Wilson Fermions}
\label{subsec:EShifts}

To investigate the effect of the nonperturbatively improved clover
coefficient in the full QCD calculations, the lowest-lying
neutral-pion background-field energy is considered
\begin{align}
  E_\piz^2(B) = \mpiz^2 - 4\pi\,\mpiz\,\beta_\piz\,B^2 + \order{B^3}\,,
\end{align}
To determine how the known additive quark mass renormalisation of
\eqnr{eqn:mqB} effects the pion mass in full QCD, we commence with the
consideration of the Gell-Mann-Oakes-Renner relation~\cite{GellMann:1968rz,Gattringer:2010zz}
\begin{align}
  \mpiz^2 &= -\frac{2\,m_{u/d}}{f_\pi^2}\,\braket{\Omega|\ub u | \Omega} \nonumber \\
  &\equiv m_{u/d}\,E_\Omega\,.
\end{align}
where $E_\Omega = -2\,\braket{\Omega|\ub u | \Omega}/f_\pi^2$, $f_\pi$
is the pion decay constant and $\braket{\Omega|\ub u | \Omega}$ is the
chiral condensate.  As $m_\pi^2 \propto m_q$, $E_\Omega$ has a
relatively weak quark mass dependence.

Using \eqnr{eqn:mqB}, we introduce a background-field dependent pion
mass due to the Wilson additive mass renormalisation
\begin{align}
  m_{[\piz]}^2 (B) &= m_{[u/d]}(B)\,E_\Omega(B) \nonumber \\
  &= \rb{m_{u/d} + \frac{a\, \xi}{2}\,\abs{q_{u/d}e B}}\,E_\Omega(B)\,.
  \label{eqn:GORW}
\end{align}
Because QCD effects will modify \eqnr{eqn:mqB}, we have introduced a
coefficient $\xi$ in \eqnr{eqn:GORW} which in principle can be
$B$-field dependent.

For example, Bali {\it et al.}~\cite{Bali:2017ian} investigated the
change in the quark mass in Wilson-fermion QCD plus background
magnetic-field simulations by examining the change in the critical
hopping parameter as a function of magnetic field strength.  For small
external magnetic-field strengths the mass shift is an order of
magnitude smaller from the free-field case and for their smallest
field strengths, the sign of the shift differs.  The dependence of
\eqnr{eqn:mqB} begins to emerge at large magnetic-field strengths as
QCD effects become small.  As discussed in the following subsection,
our nonperturbatively improved clover-fermion results also display
this order of magnitude suppression.  However our survey of
magnetic-field strengths does not reveal the Wilson-fermion sign
change in the mass shift.  Thus it is sufficient to treat $\xi$ as a
constant to be determined.

Studies~\cite{DElia:2011koc,Bali:2011qj} have  indicated that $E_\Omega$ changes
slowly in an external magnetic field and on this basis, we consider the
leading order approximation
\begin{align}
  m_{[\piz]}^2(B) &\simeq \rb{m_{u/d} + \frac{a\,\xi}{2}\,\abs{q_{u/d}e B}}\,E_\Omega\rb{0} \nonumber \\
  &\simeq \mpiz^2 + \frac{a\,\xi}{2}\,\abs{q_{u/d}e B}\,E_\Omega\rb{0}\,.
  \label{eqn:supWmpiz}
\end{align}
Finally the energy of a neutral pion in an external magnetic field using Wilson fermions is
\begin{align}
  E_{[\piz]}^2(B) &= m_{[\piz]}^2 - 4\pi\,m_{[\piz]}\,\beta_\piz\,B^2 + \order{B^3}, \nonumber\\
                     &\simeq \mpiz^2 + \frac{a\,\xi}{2}\,E_\Omega\rb{0}\,\abs{q_{u/d}e B} \nonumber \\
  &\quad - 4\pi\,\beta_\piz\,B^2\,\sqrt{\mpiz^2 +
    \frac{a\,\xi}{2}\,E_\Omega\rb{0}\,\abs{q_{u/d}e B}}\, .
  \label{eqn:WEpiN} 
\end{align}
We note that terms linear in $B$ from the magnetic field dependence of
$E_\Omega$ in \eqnr{eqn:GORW} can combine with the linear term
$\frac{a\,\xi}{2}\,\abs{q_{u/d}e B}$ to provide a contribution
proportional to $B^2$, thus contaminating the $\order{B^2}$ signal
used to extract $\beta_\piz$.  Ultimately, it is important to ensure
this ${\cal O}(a)$ term is removed.

To explore the presence of additive quark-mass renormalisation, we
focus on the quantity
\begin{align}
  E_{[\piz]}^2(B) -& \mpiz^2
                     \simeq \frac{a\,\xi}{2}\,E_\Omega\rb{0}\,\abs{q_{u/d}e B} \nonumber \\
  &\ \ - 4\pi\,\beta_\piz\,B^2\,\sqrt{\mpiz^2 +
    \frac{a\,\xi}{2}\,E_\Omega\rb{0}\,\abs{q_{u/d}e B}} \, .
  \label{eqn:E2submpiz2}
\end{align}
This energy shift can be constructed using the two correlator combinations
\begin{align}
  R_+(B,t) &= G(B,t)\,G(0,t)\, , \label{eqn:R+} \\
  R_-(B,t) &= \frac{G(B,t)}{G(0,t)}\, , \label{eqn:R-}
\end{align}
where $G(B,t)$ is the zero-momentum projected two-point correlation
function. Upon taking the effective energy
\begin{align}
  E_{\text{eff}}\rb{t} = \frac{1}{\delta\,t}\,\log\rb{\frac{G(t)}{G\rb{t + \delta\,t}}}\, ,
  \label{eqn:ee}
\end{align} 
\eqnr{eqn:R+} yields $\left( E_{[\piz]}(B) + \mpiz\right)$ and
\eqnr{eqn:R-} provides $\left( E_{[\piz]}(B) - \mpiz\right)$. These
effective energy shifts from $R_+$ and $R_-$ are then multiplied
together to form the $E_{[\piz]}^2(B) - \mpiz^2$ energy shift of
\eqnr{eqn:E2submpiz2}. We note that \eqnr{eqn:R-} is particularly helpful in isolating $B$-depdendent terms, as QCD contributions are correlated in the ratio of correlation functions and largely cancel.

Noting \eqnr{eqn:E2submpiz2} has leading linear and quadratic terms in
$B$ we commence by considering the fit function
\begin{align}
  E^2_{[\piz]}(B) -\mpiz^2 = c_1\,k_B + c_2\,k_B^2 \, .
  \label{eqn:c1c2}
\end{align}
Here $k_B$ is the quantisation number from the quantisation condition on
magnetic field strength of \eqnr{eqn:eB}.  An estimate for the fit
parameter $c_1$ can be obtained from Eq.~(\ref{eqn:E2submpiz2})
\begin{align}
  c_1 = \,\frac{\pi}{a}\,\abs{\frac{q_{u/d}}{q_d}}\,\frac{\xi\,E_\Omega\rb{0}}{N_x\,N_y}\, .
  \label{eqn:c1-1}
\end{align}
Recalling the Gell-Mann-Oakes-Renner relation at zero magnetic field
provides $E_\Omega\rb{0} = \mpiz^2 / m_{q}$ and drawing on the Wilson
quark-mass relation
\begin{align}
  m_q = \frac{1}{2\,a}\,\rb{\frac{1}{\kappa} - \frac{1}{\kappa_{cr}}}\,,
\end{align}
where $\kappa_{cr}$ is the critical hopping parameter where the
zero-field pion mass vanishes, Eq.~(\ref{eqn:c1-1}) can be written
\begin{align}
  c_1 = 2\pi\,\abs{\frac{q_{u/d}}{q_d}}\,\frac{\xi\,\mpiz^2}{N_x\,N_y}\,\rb{\frac{1}{\kappa} - \frac{1}{\kappa_{cr}}}^{-1}\,.
  \label{eqn:c1}
\end{align}
Similarly, drawing on Eq.~(\ref{eqn:E2submpiz2}), $c_2$ is related to the magnetic polarisability 
\begin{align}
  \beta = -c_2 \,\alpha\,\frac{q_d^2\,a^4}{m_{[\piz]}}\,\left(\frac{N_x\,N_y}{2\pi}\right)^2\,,
  \label{eqn:c2}
\end{align}
where $m_{[\piz]}$ is provided in \eqnr{eqn:supWmpiz} and $\alpha=
1/137.036$ is the fine structure constant.  Of course, if $c_1
\neq 0$ then the magnetic field dependence of $E_\Omega$ in
Eq.~(\ref{eqn:c1-1}) will induce additional ${\cal O}(B^2)$
contaminations to Eq.~(\ref{eqn:c2}).  Thus it is vital to remove this
${\cal O}(a)$ error.

As the $u$ and $d$ quarks are mass degenerate in our lattice QCD
simulation, the neutral pion correlation functions with quark content
$u\ub$ or $d\db$ differ only by the strength of the quark flavour
interactions with the external magnetic field. This allows the neutral
pion correlation function to be evaluated at a larger range of field
strengths than is possible for the $\pi^+$, using the same
propagators.

\subsection{Nonperturbatively Improved Clover Fermions}
\label{sec:NPICF}

\begin{figure*}[t!]
\includegraphics[width=\columnwidth]{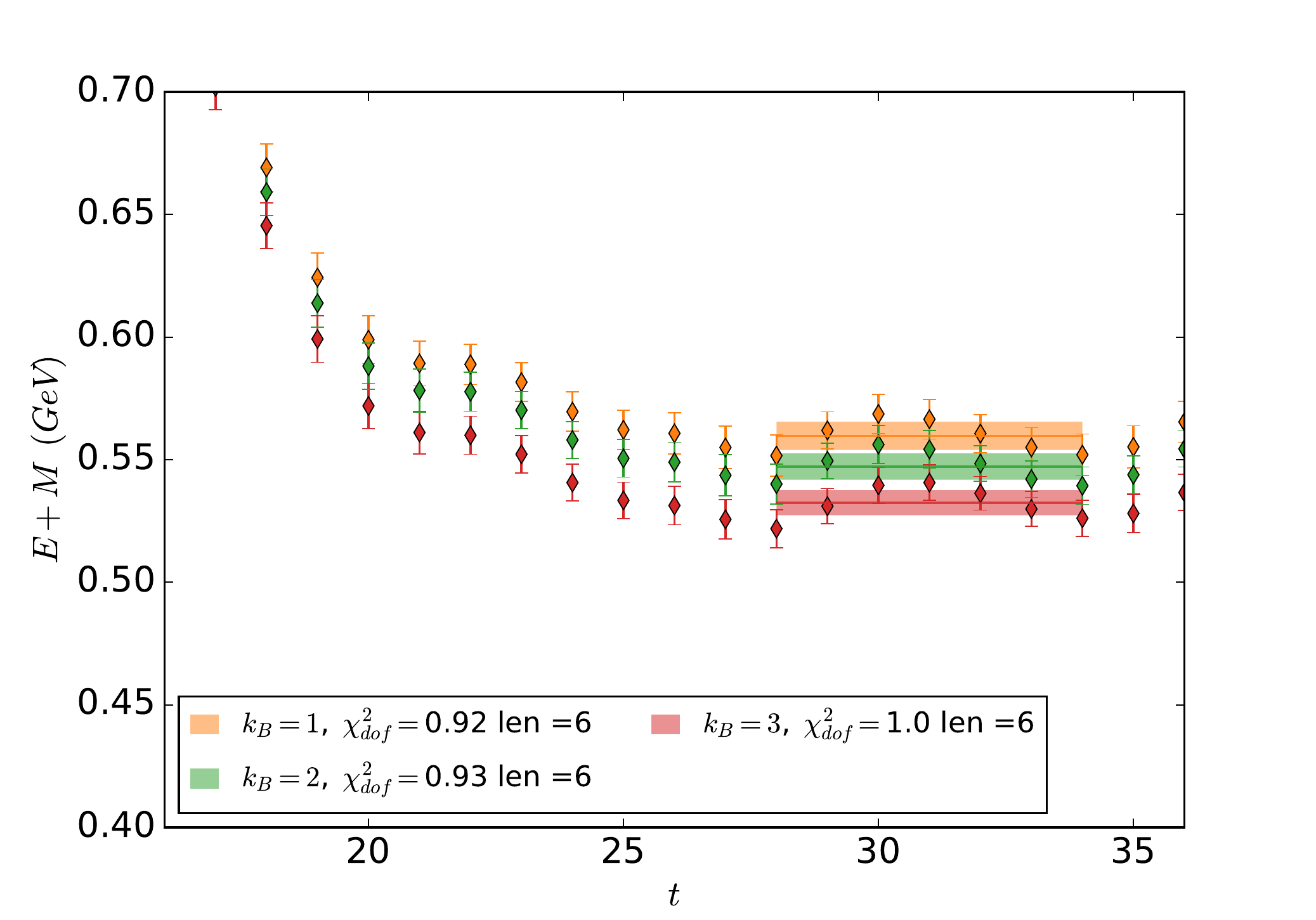}
\includegraphics[width=\columnwidth]{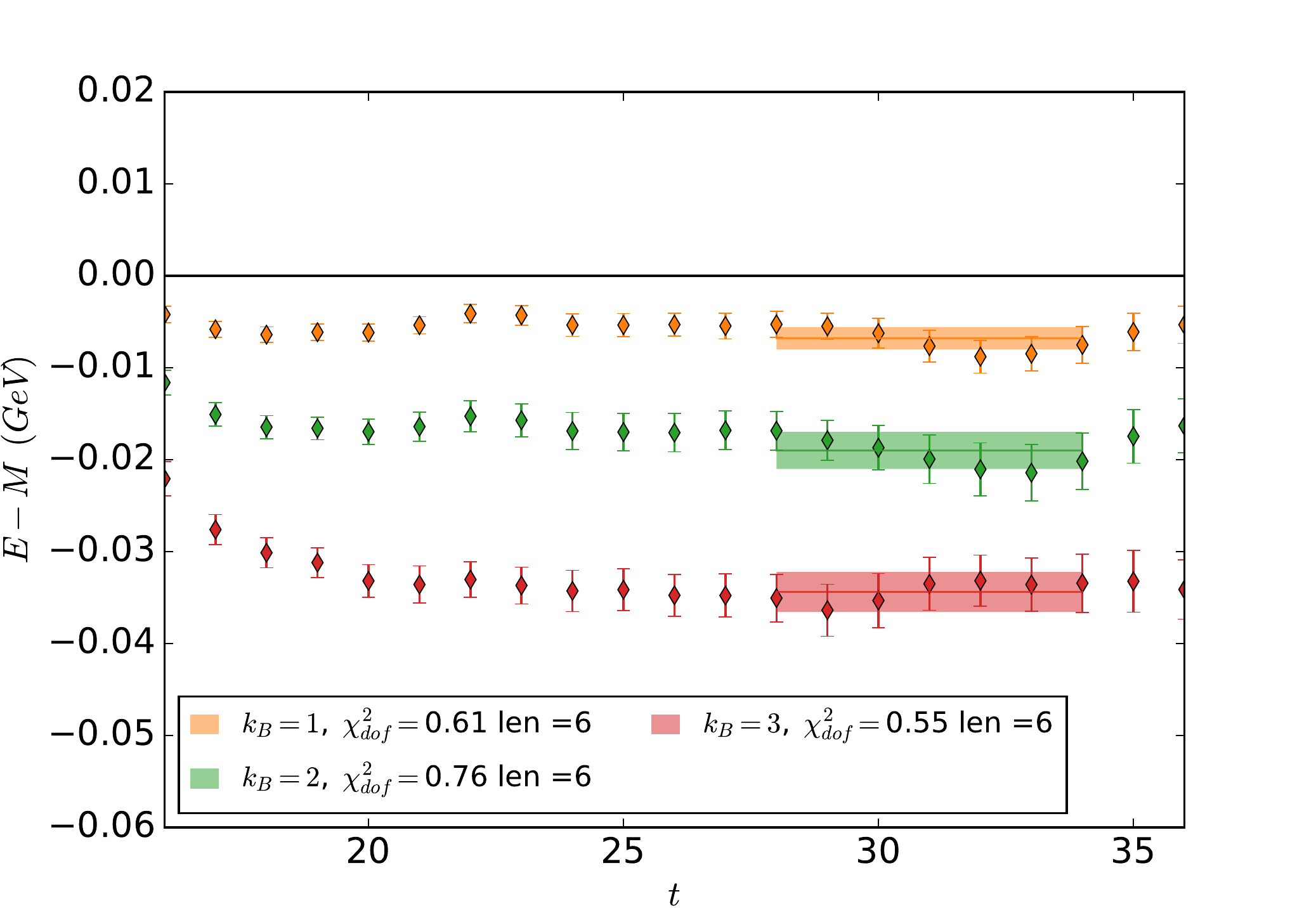}
	\caption{Neutral pion energy shifts from
          \eqnrtwo{eqn:R+}{eqn:R-} for $E_{[\piz]}(B) + \mpiz$
          (left) and $E_{[\piz]}(B) - \mpiz$ (right) respectively
          using a nonperturbatively improved clover fermion action on
          the $m_\pi = 296$ MeV ensemble.  The three smallest field
          strengths are illustrated.  Shaded regions illustrate the
          fit windows selected through the consideration of the full
          covariance-matrix $\chi^2_{dof}$, the extent of the fit
          window and the desire to select the same fit window for all
          effective-energy shifts.}
	\label{fig:piuuddEfitsplots_MixKappa}
\end{figure*}

\begin{figure}[h!]
  \includegraphics[width=\columnwidth]{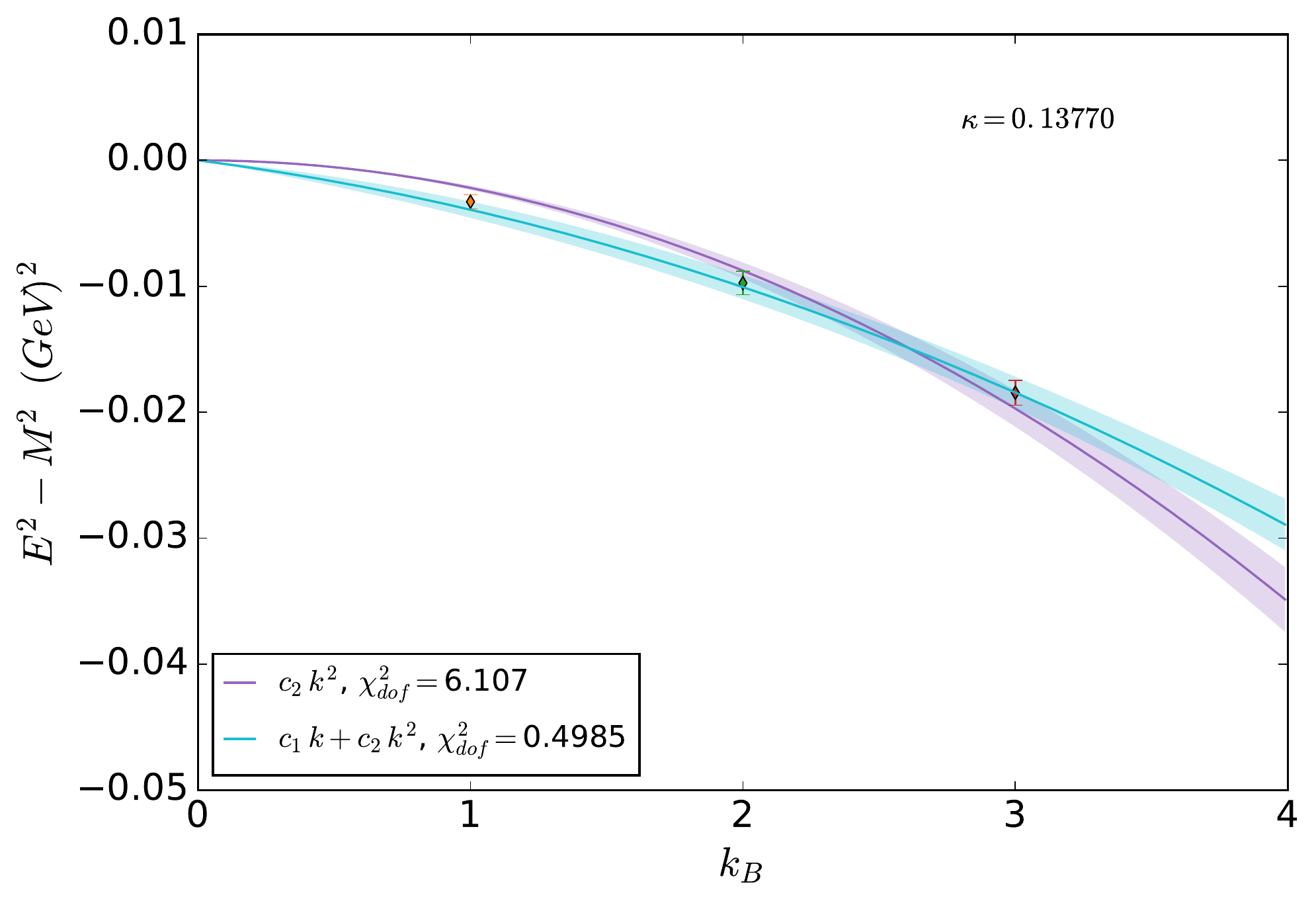}
  \caption{Fits of the magnetic-field induced energy shift to
    the magnetic-field quanta for the nonperturbatively improved
    clover fermion action.  The full covariance-matrix based
    $\chi^2_{dof}$ provides evidence of a nontrivial value for
    fit coefficient $c_1$, indicating the presence of unwanted
    Wilson-like additive mass renormalisations in the
    nonperturbatively improved clover fermion action.}
  \label{fig:MixKappa_Eshiftfits}
\end{figure}

The nonperturbatively improved clover fermion action used in the full
QCD calculations differs from that used in the free-field calculations
by the value of the clover coefficient $c_{\textrm{sw}}$ multiplying the ${\cal
  O}(a)$ clover term of the fermion action in Eq.~(\ref{eqn:clover}).

%
The free-field
simulations used the tree-level value $c_{\textrm{sw}} = 1$ while the full
QCD calculations use the nonperturbatively improved value $c_{\textrm{sw}} =
1.715$~\cite{Aoki:2005et,Aoki:2008sm}. The extent to which this
changes the cancellation of the additive mass renormalisation seen to
occur in \Fig{fig:FF_Clover_action} is investigated using the energy
shift defined above in \eqnr{eqn:E2submpiz2}.

The $E_{[\piz]}(B) - \mpiz$ energy shift is illustrated in
\Fig{fig:piuuddEfitsplots_MixKappa} right; it is quite clear that it
is easy to construct good plateau fits for this energy shift. This is
in contrast to the $E_{[\piz]}(B) + \mpiz$ energy shift in
\Fig{fig:piuuddEfitsplots_MixKappa} left, as the correlated QCD
fluctuations between field strengths compound rather than cancel,
making it difficult to fit constant plateaus. This difficulty in
fitting constant plateaus reduces the fit parameter space considered
as common plateau fits are required for both $E_{[\piz]}(B) -
\mpiz$ and $E_{[\piz]}(B) + \mpiz$. The fit window $t = \left[
  28,34\right]$ is chosen as this window has good fits with acceptable
$\chi^2_{dof}$'s across each field strength and energy-shift type
considered.

The energy shifts are fitted as a function of the field-strength
quanta $k_B$ in \Fig{fig:MixKappa_Eshiftfits}.
Recalling \eqnr{eqn:E2submpiz2} has leading linear and quadratic terms in
$B$ we consider the fit function
\begin{align}
  E^2_{[\piz]}(B) -\mpiz^2 = c_1\,k_B + c_2\,k_B^2 \, .
  \label{eqn:c1c2r}
\end{align}
Under the assumption of the removal of additive mass renormalisation,
we first consider fixing $c_1=0$ and using a $c_2\,k_B^2$ quadratic-only
fit function.  As illustrated in \Fig{fig:MixKappa_Eshiftfits} this
provides a poor description of the results and yields an unacceptable
$\chi^2_{dof} = 6.1$.

Allowing for a nontrivial $c_1$ coefficient provides the linear +
quadratic, $c_1\,k_B + c_2\,k_B^2$ fit of \eqnr{eqn:c1c2r}.  This fit
describes the lattice simulation results well with a $\chi^2_{dof} =
0.5$.  However, it also indicates the presence of additive mass
renormalisation in the nonperturbatively improved clover fermion
simulation.

To explore the necessity of this ${\cal O}(a)$ term linear in the
magnetic field strength, a quadratic + cubic, $c_2\,k_B^2 + c_3\,k_B^3$,
fit is also considered.  However, this model does not describe the
simulation results, producing unacceptably high $\chi^2_{dof}$ values.
Fits over four field strengths are also considered.  This time, the
acceptable fit requires linear, quadratic and cubic terms.  Thus, the
demand for an ${\cal O}(a)$ term associated with the presence of
additive mass renormalisation is robust.

\begin{figure*}[t!]
\includegraphics[width=\columnwidth]{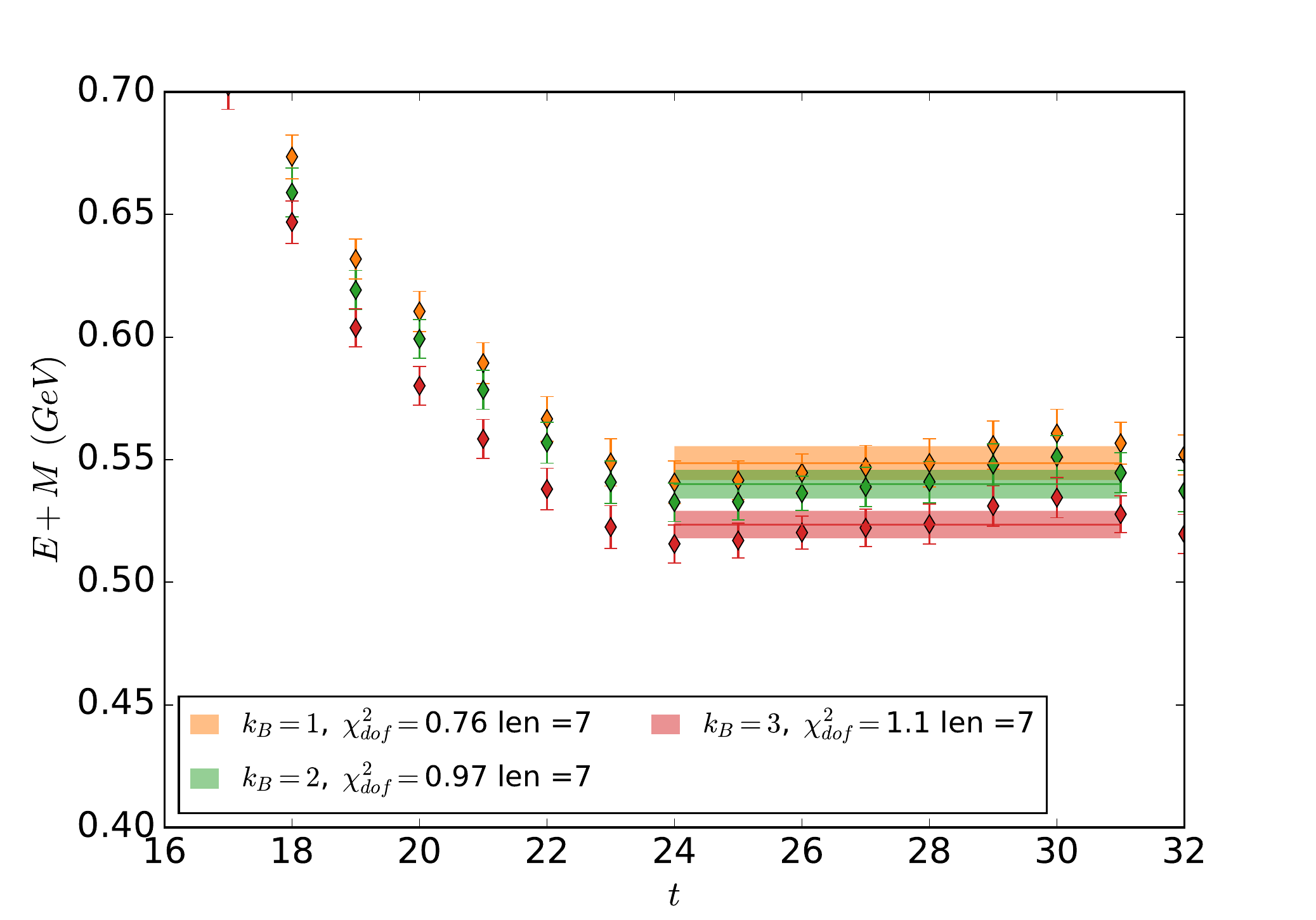}
\includegraphics[width=\columnwidth]{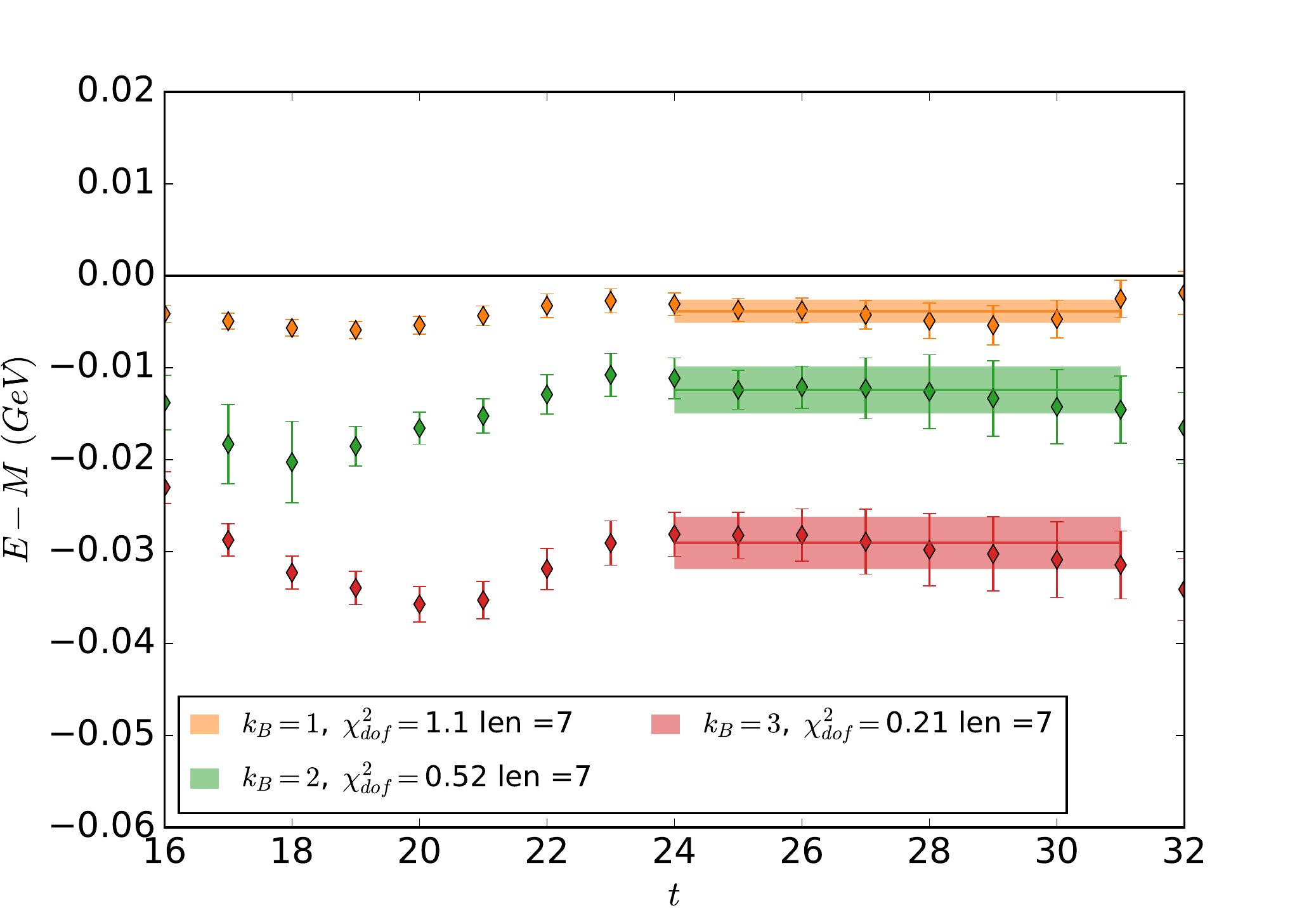}
	\caption{Neutral pion energy shifts from
          \eqnrtwo{eqn:R+}{eqn:R-} for $E_\piz(B) + \mpiz$ (left) and
          $E_\piz(B) - \mpiz$ (right) respectively using the BFC clover fermion
          action on the $m_\pi = 296$ MeV ensemble.  The three
          smallest field strengths are illustrated.  Shaded regions
          illustrate the fit windows selected through the
          consideration of the full covariance-matrix $\chi^2_{dof}$,
          the extent of the fit window and the desire to select the
          same fit window for all effective-energy shifts.}
	\label{fig:piudd_Efitsplot_EMClover}
\end{figure*}

\subsection{Expected Mass Renormalisation}

These results indicate that the current treatment of the clover term
is inadequate.  In light of the success observed in the free field
limit, we anticipate that the application of the nonperturbatively improved
value $c_{\textrm{sw}}^{\textrm{NP}} = 1.715$ to both the QCD and the background field
contributions to the clover term has spoiled the removal of additive
mass renormalisation associated with the magnetic field.

If the tree-level value of $c_{\textrm{sw}}^{\textrm{tree}} = 1$ is the value
required for the removal of ${\cal O}(a)$ errors, then we have over
compensated by an amount
\begin{align}
D^{\rm NP} = c_{\textrm{sw}}^{\rm tree} - c_{\textrm{sw}}^{\rm NP} = 1.0 - 1.715 = -0.715 \, .
\end{align}
This over compensation is relative to the standard Wilson action
discrepancy of
\begin{align}
D^{\rm W} = c_{\textrm{sw}}^{\rm tree} = 1.0 \, .
\end{align}
This over compensation factor can be incorporated into the
Wilson-fermion discussion of Section~\ref{sec:NPICF}
through 
\begin{align}
  \xi \to \xi^{\rm NP} = \frac{D^{\rm NP}}{D^{\rm W}}\, \xi \, ,
  \label{eqn:xiNP}
\end{align}
enabling a prediction of the nontrivial value for $c_1$.  Including
the aforementioned order of magnitude suppression factor $\xi = 1/10$,
\eqnrtwo{eqn:c1}{eqn:xiNP} provide an estimate for the fit parameter $c_1$.  For
the $\pi^0_d$ with $m_\pi = 296$ MeV, $\kappa_{d} = 0.13770$ and
$\kappa_{cr} = 0.13791$~\cite{Aoki:2008sm}
\hbox{$c_1 \sim -3 \times 10^{-3}$ GeV$^2$}.  From the linear +
quadratic $c_1\,k_B + c_2\,k_B^2$ fit \hbox{$c_1 = -2.8(9) \times
  10^{-3}$ GeV$^2$}.

\section{Background Field Correction}
\label{sec:EMClover}

Motivated by the analytic calculation in Section~\ref{sec:FFEMClover},
and encouraged by the above result, in this section we investigate a new
clover fermion action where the coefficients for the QCD and background
magnetic field contributions to the field-strength tensor in
Eq.~(\ref{eq:emclover}) take different values. The electromagnetic
clover coefficient takes the tree level value $c_{\textrm{em}} = 1,$
whilst the QCD clover coefficient retains its non-perturbative value
of $c_{\textrm{sw}}^{\rm NP} = 1.715.$ In the following, we refer to
this modified fermion action as the Background Field Corrected (BFC)
form of the clover fermion action.

\subsection{Additive Mass Renormalisation}
\label{sec:EMCloverAMR}

The process of calculating correlation functions and forming energy
shifts is repeated as detailed above in Eqs.~(\ref{eqn:R+}) through
(\ref{eqn:ee}).  Figure~\ref{fig:piudd_Efitsplot_EMClover} displays
the new effective energy shifts and the associated fits.

In the absence of additive mass renormalisations,
\eqnr{eqn:E2submpiz2} simplifies to
\begin{align}
  E_\piz^2(B) -& \mpiz^2 = - 4\pi\,\mpiz\,\beta_\piz\,B^2\, + {\cal O}(B^3) \, ,
  \label{eqn:E2submpiz2clean}
\end{align}
and therefore we consider fit functions of the form
\begin{align}
  E^2_{\piz}(B) -\mpiz^2 = c_1\,k_B + c_2\,k_B^2 + c_3\, k_B^3 \, .
  \label{eqn:c1c2c3}
\end{align}
If the modified clover action has removed the Wilson additive mass
renormalisation, the linear term of \eqnr{eqn:c1c2c3} will have a
trivial coefficient.  Indeed, fits without the linear term should
describe the energy shifts well.

\begin{figure}[t!]
	\includegraphics[width=\columnwidth]{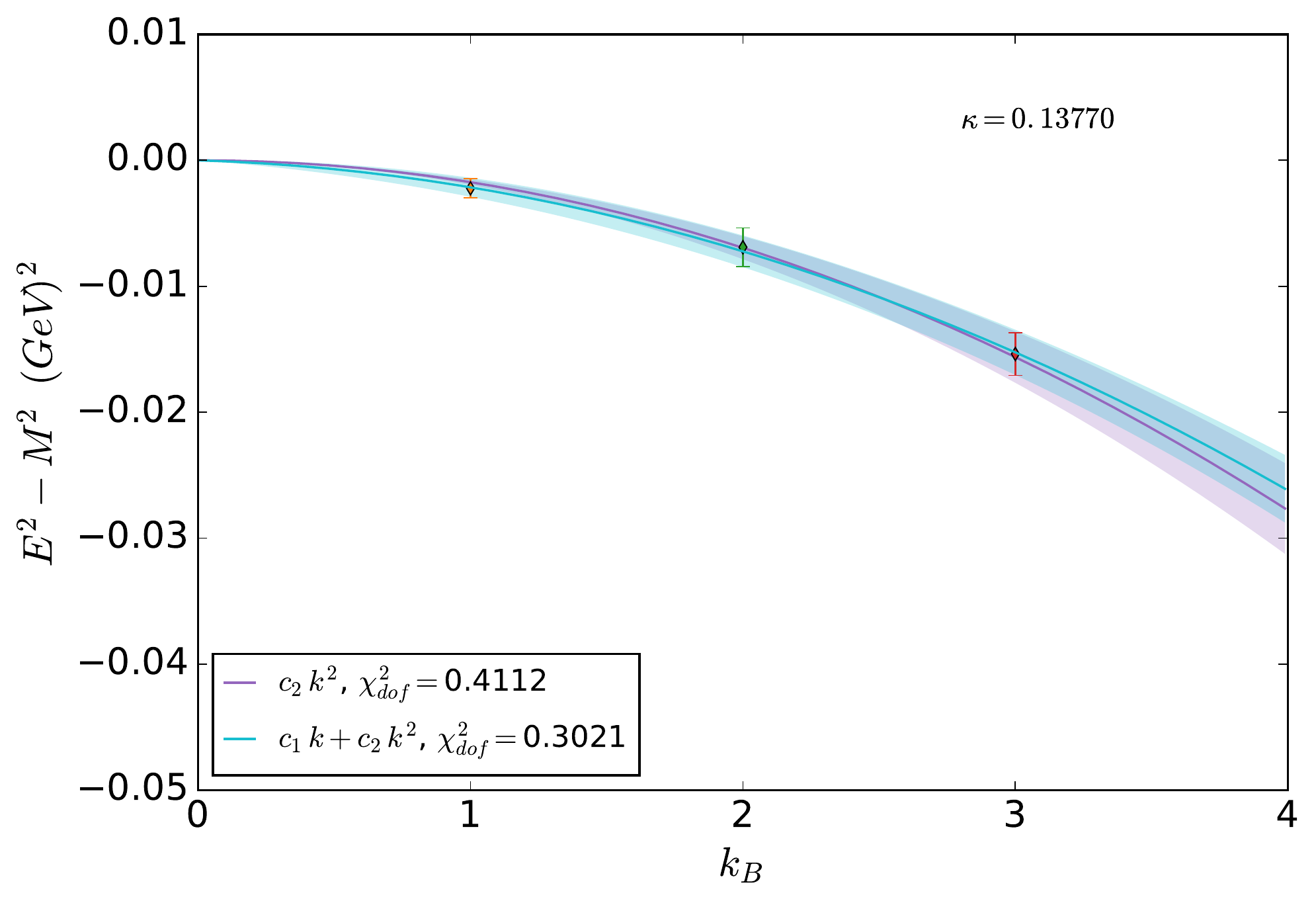}
	\caption{Fits of the magnetic-field induced energy shift to
          the magnetic-field quanta for the BFC clover action.
          The full covariance-matrix based $\chi^2_{dof}$ for the
          simple $c_2\, k_B^2$ quadratic fit provides evidence of the
          elimination of Wilson-like additive mass renormalisations in
          the BFC clover action. Allowing for a non-trivial
          value of $c_1$ produces a value consistent with zero.}
	\label{fig:pid_Efitsplots_EMClover}
\end{figure}

Fits of \eqnr{eqn:c1c2c3} to results from the lowest three magnetic
field strengths are illustrated in \Fig{fig:pid_Efitsplots_EMClover}.
As acceptable fits are obtained with the two leading terms, $c_3$ has
been constrained to zero.  The success of our BFC clover fermion action is
reflected in the excellent description of the results using only a
quadratic term in \eqnr{eqn:c1c2c3}, in accord with the expectations
of \eqnr{eqn:E2submpiz2clean}.  Allowing for the possibility of a
nontrivial value for $c_1$, we find $c_1 = (-5.3\pm 8.7)\times
10^{-4}\ {\rm GeV}^2$, consistent with zero.

\begin{figure}[t!]
	\includegraphics[width=\columnwidth]{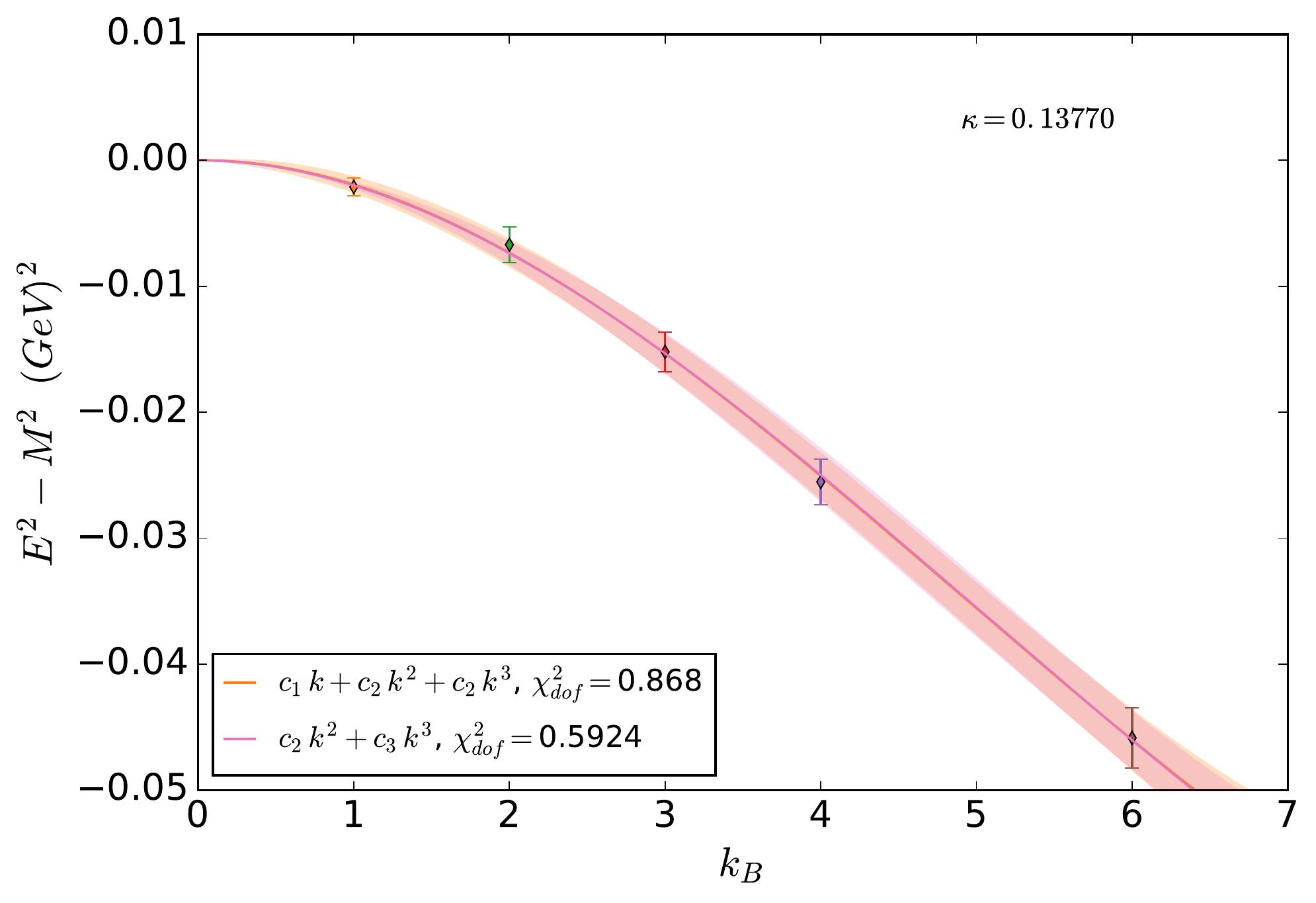}
	\caption{Fits of the magnetic-field induced energy shift to
          the first five magnetic-field quanta for the BFC clover action.
          Here a term cubic in the magnetic field strength is required
          to describe the largest field strength.}
	\label{fig:piuudd_Efitsplots_EMClover}
\end{figure}

Drawing on the full range of five magnetic field strengths available
to $d \overline d$ and $u \overline u$ pions,
\Fig{fig:piuudd_Efitsplots_EMClover} illustrates fits of
\eqnr{eqn:c1c2c3} to the BFC clover results.  This time the
results demand $c_3 \ne 0$ and thus a $B^3$ term is manifest.
Again, the success of our BFC clover action in removing additive mass
renormalisations is reflected in the excellent description of the
results using only quadratic and cubic terms in \eqnr{eqn:c1c2c3}, in
accord with the expectations of \eqnr{eqn:E2submpiz2clean}.  Allowing
for the possibility of a nontrivial value for $c_1$, we find $c_1 =
(1.1\pm 9.0)\times 10^{-4}\ {\rm GeV}^2$, consistent with zero.

We note that similar results are observed for fits to the first four
magnetic field strengths.  The lattice results are described well by
fits with $c_1$ constrained to zero, and allowing for a nontrivial
value provides $c_1 = (-6.2\pm 8.2) \times 10^{-4}\ {\rm GeV}^2$,
again consistent with zero.  While it is sufficient to set $c_3 = 0$,
the onset of nontrivial $\order{B^3}$ terms lies between the fourth
and fifth field strengths and therefore, only the three smallest field
strengths are considered in determining the magnetic polarisability.

In summary, the BFC clover fermion action has successfully removed the
additive quark mass renormalisation due to the Wilson term.  The key
is to employ the tree-level value $c_{\textrm{em}} = 1$ for uniform
background field contributions to the electromagnetic clover term.

\subsection{Magnetic Polarisability}
\label{sec:MagPol}

Having demonstrated the removal of additive quark-mass renormalisation
in the BFC clover action, the magnetic polarisability can be
determined without concern for the aforementioned ${\cal O}(a)$
contaminations entering the $B^2$ associated with the magnetic
polarisability.

The energy shift and fits performed in Section~\ref{sec:EMCloverAMR}
are used and the polarisability extracted from the coefficient $c_2$
of the quadratic term of \eqnr{eqn:c1c2c3}.  Comparing with
\eqnr{eqn:E2submpiz2clean} and drawing on the field quantisation
condition of \eqnr{eqn:qc} with $q = q_d = -1/3$, the magnetic polarisability
is given by
\begin{align}
  \beta = -c_2 \,\alpha\,\frac{q_d^2\,a^4}{m_\pi}\,\left(\frac{N_x\,N_y}{2\pi}\right)^2
        \, .
\end{align}

\begin{table}[t]
\caption{Magnetic polarisability of the neutral pion from the
  $\order{a}$-improved BFC clover action analysis of the lowest
  three magnetic-field strengths.  }
\label{tab:magPolar}
\begin{ruledtabular}
\begin{tabular}{ccc}
  $\kappa$ &$m_\pi$ (MeV) &$\beta_\piz$ ($\times 10^{-4}\ \mbox{fm}^3$) \\
  \noalign{\smallskip}
  \hline
  \noalign{\smallskip}
  0.13754  & 411          & 0.62(4) \\
  0.13770  & 296          & 0.54(7) \\
\end{tabular}
\end{ruledtabular}
\end{table}

Results for the magnetic polarisability of the neutral pion from the
$\order{a}$-improved BFC clover action analysis of the lowest
three magnetic-field strengths are reported in Table~\ref{tab:magPolar}.  While the report of our analysis has focused on
the light-quark hopping parameter of $\kappa_{ud} = 0.13770$
corresponding to $296$ MeV, results for $m_\pi = 411$ are also
reported.

As discussed in the previous section, we consider the first three
field strengths to avoid the possibility of complications associated
with non-trivial $\order{B^3}$ contributions.  While the inclusion of
a fourth point takes the magnetic polarisability from $\beta_\piz =
0.54(7)$ to $0.52(5) \times 10^{-4}\ \mbox{fm}^3$, the small
improvement in statistical uncertainty may be offset by an increase in
systematic uncertainty associated with non-trivial $\order{B^3}$
contributions.

\section{Conclusions}
\label{sec:Conclusions}

In this paper the response of the pion to a uniform background
magnetic field has been investigated. The existence of a
field-strength dependent additive quark-mass renormalisation
associated with Wilson fermions was confirmed. We performed an analytic
calculation showing that at tree-level in the QCD-free case the clover
term corrects for these spurious contributions.

When QCD interactions are included, a careful treatment of the QCD and
electromagnetic clover terms in Eq.~(\ref{eq:emclover}) is required. While
the interactions of QCD require the nonperturbatively improved value
of $c_{\textrm{sw}} = 1.715$, it is essential to apply the tree-level
value $c_{\textrm{sw}}=1$ to the background magnetic field
contributions.  With this treatment, it is possible to simultaneously
remove the $\order{a}$ errors in the QCD contributions, and correct
for the $\order{a}$ errors induced by the Wilson term in a background magnetic
field.  We refer to this modified fermion action as the Background
Field Corrected clover fermion action.

With the suppression of $\order{a}$ errors associated with additive
mass renormalisation in the BFC clover action, the magnetic
polarisability of the neutral pion can now been determined.  For the
first time, a fully relativistic approach to the energy shift is used.
Results are summarised in Table~\ref{tab:magPolar} and we anticipate
these results will facilitate consensus within the field.
$\beta_\piz$ is positive, such that the energy of a pion in a magnetic
field decreases.  The extent to which this trend continues is of interest
and our results give a hint that the $\order{B^3}$ contributions
soften this trend.

Future work will approach the physical quark mass regime and interface
with chiral perturbation theory studies~\cite{Bellucci:1994eb}.
It will also be important to directly include sea-quark-loop
interactions with the background magnetic field to incorporate these
contributions to the magnetic polarisability.  However, this approach
is prohibitively expensive.  It requires a separate Monte-Carlo
ensemble for each field strength considered and this will result in a
loss of important QCD correlations between different magnetic field
strengths.

An alternative approach is to separate the valence and sea-quark-loop
contributions to the magnetic polarisability in effective field
theory~\cite{Hall:2013dva,Bignell:2018acn}.  This will enable an accurate extrapolation of current lattice
QCD results and an estimation of sea-quark-loop contributions to the
magnetic polarisabilities of the pion.

\section*{Acknowledgements}

We thank the PACS-CS Collaboration for making their $2+1$ flavour
configurations available and the ongoing support of the International
Lattice Data Grid (ILDG). This work was supported with supercomputing
resources provided by the Phoenix HPC service at the University of
Adelaide. This research was undertaken with the assistance of
resources from the National Computational Infrastructure (NCI). NCI
resources were provided through the National Computational Merit
Allocation Scheme, supported by the Australian Government through
Grants No.~LE190100021, LE160100051 and the University of Adelaide
Partner Share.  R.B. was supported by an Australian Government
Research Training Program Scholarship.  This research is supported by
the Australian Research Council through Grants No.~DP140103067,
DP150103164, DP190102215 (D.B.L) and DP190100297 (W.K).
\newpage

\bibliography{pbase}

\end{document}